\documentclass[a4paper]{jpconf}
\usepackage{graphicx}
\begin{document}
\title{Pulsating B-type stars in the young open cluster\\
h Persei (NGC 869)}

\author{Aleksandra Majewska-\'Swierzbinowicz$^1$, Andrzej Pigulski$^1$,\\ R\'obert Szab\'o$^{2,3}$ and Zolt\'an Csubry$^2$}

\address{$^1$ Instytut Astronomiczny Uniwersytetu Wroc{\l}awskiego, Wroc{\l}aw, Poland}
\address{$^2$ Konkoly Observatory of the Hungarian Academy of Sciences, Budapest, Hungary}
\address{$^3$ Physics Department, University of Florida, Gainesville, USA}

\ead{majewska@astro.uni.wroc.pl}

\begin{abstract}
We announce the discovery of six $\beta$~Cephei stars and many other variable stars in the young open cluster
h Persei (NGC 869). The cluster seems to be very rich in variable B-type stars, similarly to its twin, $\chi$~Persei (NGC\,884).
\end{abstract}

\section{Introduction}
The young open clusters h and $\chi$ Persei (NGC 869 and 884) form a well-known bright double cluster in Perseus. 
The clusters have similar age, distance and contain comparable number of stars. Recently, $\chi$~Persei was the 
target of a large international campaign aimed at the discovery and subsequent asteroseismology of pulsating stars, 
in particular $\beta$ Cephei stars \cite{vienna,sophie}. Preliminary reduction led to the 
discovery of seven $\beta$ Cephei stars in this cluster which makes it one of the richest in this kind of variable.
It can be suspected that h Persei, the twin of $\chi$~Persei, also contains a large number of early-type pulsating 
stars. For this reason, we reduced and analyzed CCD observations of h~Persei obtained during 8 observing nights 
in 2001 at Piszk\'estet\H{o} station of the Konkoly Observatory (Hungary). We present here the preliminary results 
of this study.

\section{Observations and reductions}
\begin{figure}
\begin{center}
\includegraphics[width=5in]{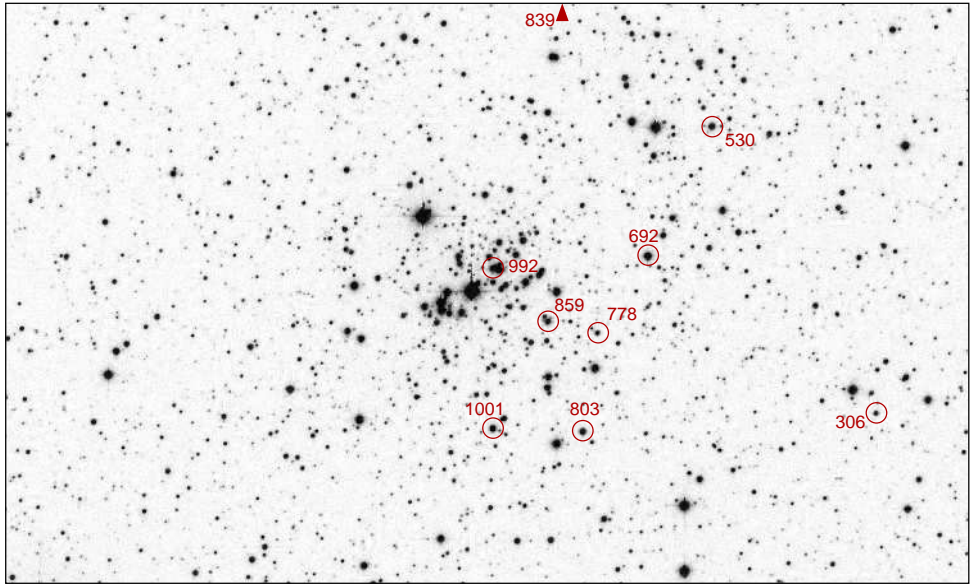}
\end{center}
\caption{\label{fig1}A sample frame (27$^\prime$$\times$18$^\prime$) of the field in h Persei (NGC 869) we observed
at Piszk\'estet\H{o}. $\beta$ Cephei stars are encircled and labeled with Oosterhoff \cite{oost37} numbers. Oo 839 is slightly outside 
this frame, although it was observed on some nights. North is up, east to the left.}
\end{figure}

The CCD data of h Persei were obtained during 8 observing nights between September 27 and October 30, 2001, 
using 60/90-cm Schmidt camera at Piszk\'estet\H{o} (Hungary). The observed field was approximately 
27$^\prime$$\times$18$^\prime$ large (Fig.~\ref{fig1}). In total, about 1600 $V$-filter CCD frames were 
acquired. The exposure times were 50-80 seconds long and, in order to avoid saturation, the telescope was 
slightly defocused. Several $I$-filter frames were taken as well. The data were calibrated in a standard way and 
then reduced by means of the Daophot package \cite{stet87}.

\begin{figure}
\includegraphics[width=20.5pc]{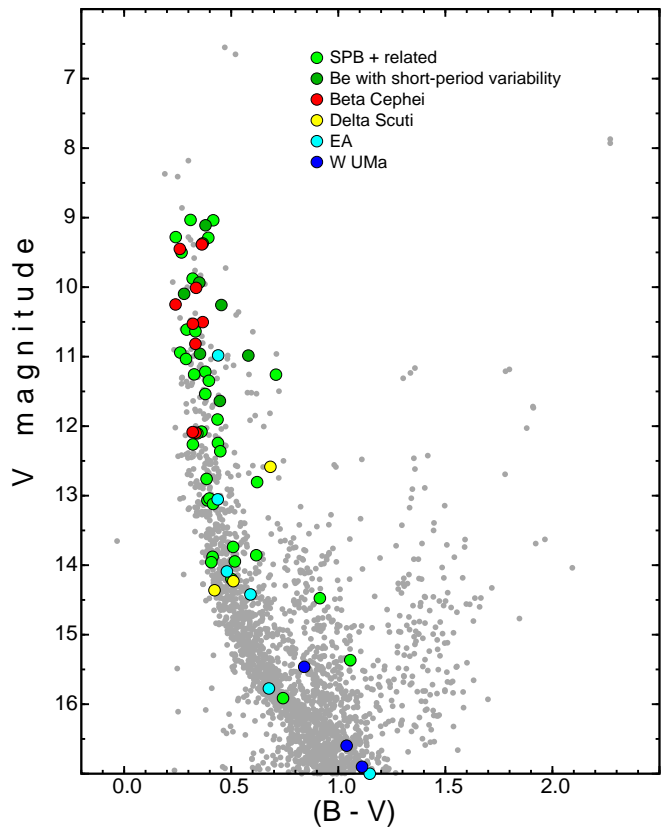}\hspace{2pc}%
\begin{minipage}[b]{11.5pc}\caption{\label{cmd}The $V$ vs.~($B-V$) co\-lour-mag\-ni\-tude diag\-ram for h Persei. The 
$BV$ pho\-to\-met\-ry was taken from \cite{kell01}. Variable stars are shown with different colours 
(la\-be\-led in the figure).}
\end{minipage}
\end{figure}

\begin{figure}
\begin{center}
\includegraphics[width=5in]{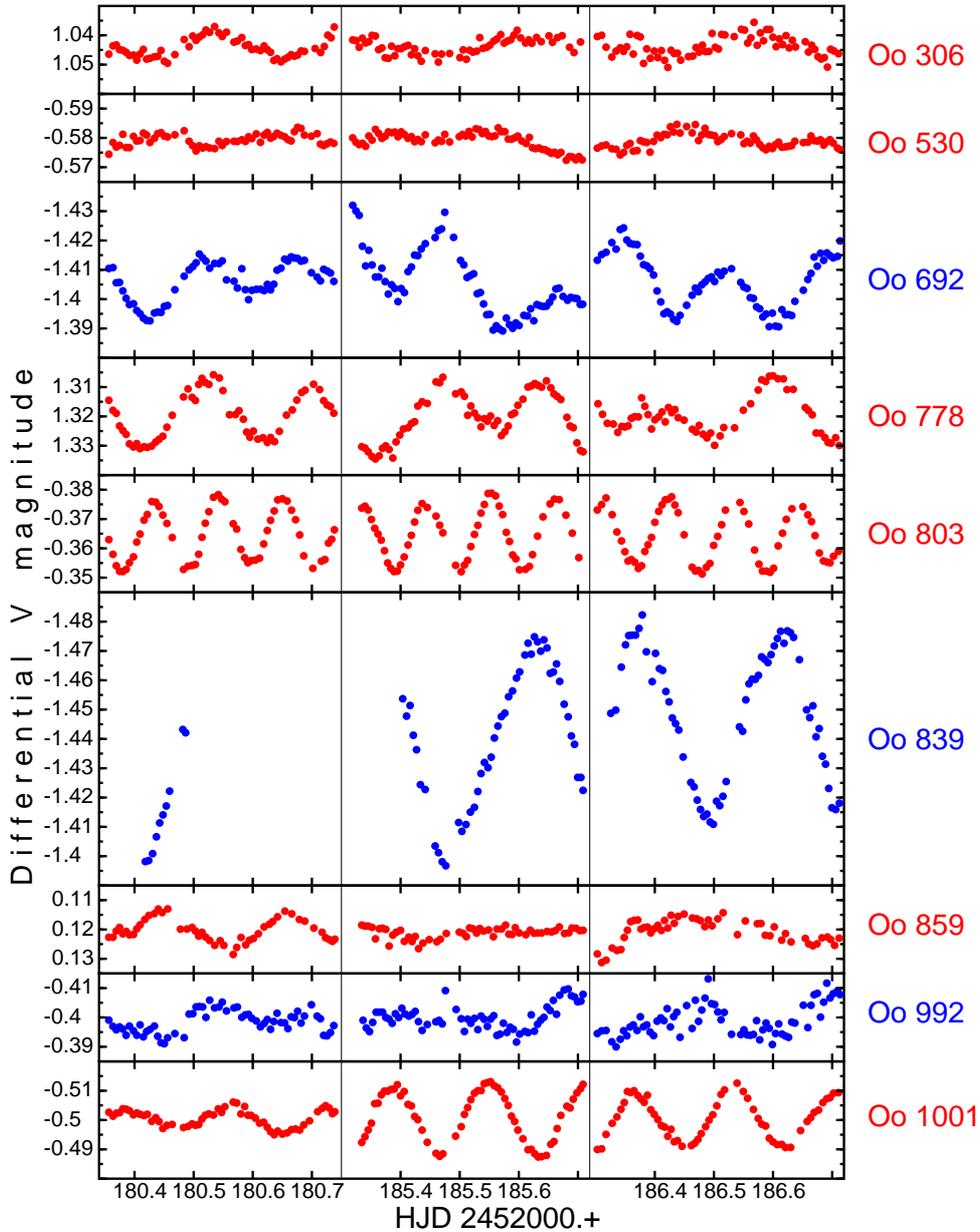}
\end{center}
\caption{\label{BCeps}Light curves on three observing nights for nine $\beta$~Cephei stars in h Persei. 
New variable stars are shown in red, already known, in blue.}
\end{figure}

\begin{figure}
\begin{center}
\includegraphics[width=5in]{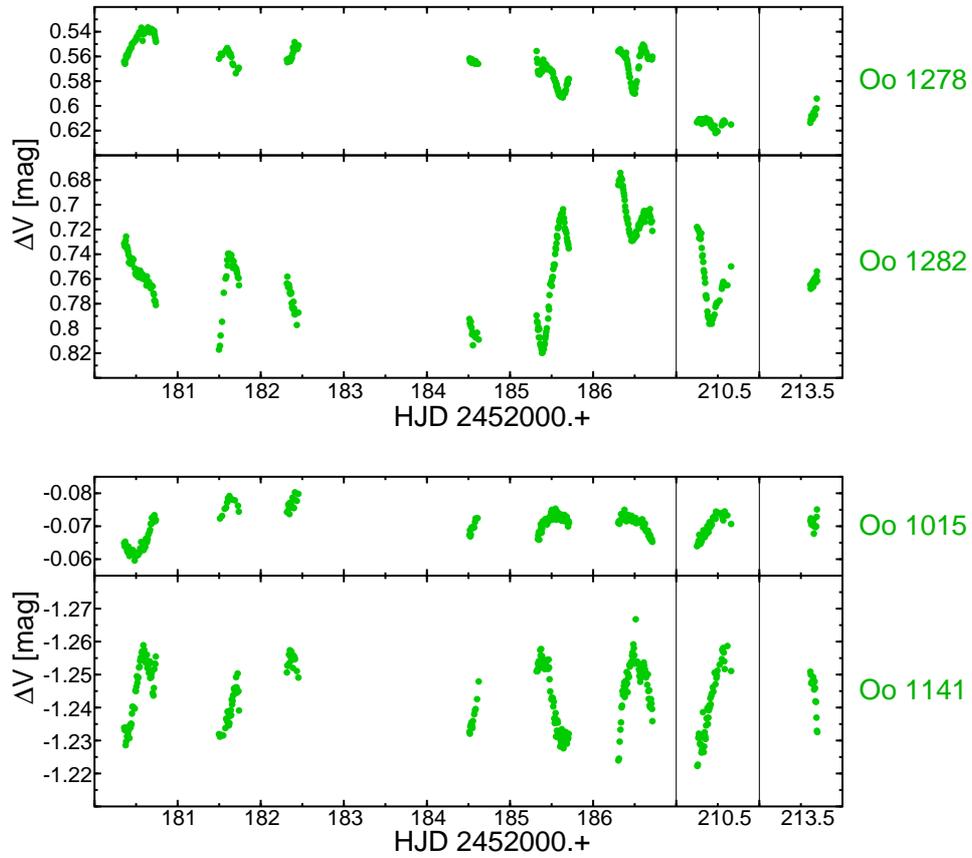}
\end{center}
\caption{\label{SPBe}Top: Light curves of two Be stars showing short-period variability, Oo 1278 and Oo 1282. 
Bottom: The same for two SPB candidates in h Persei, Oo 1015 and 1141.}
\end{figure}

\begin{figure}
\begin{center}
\includegraphics[width=3.7in]{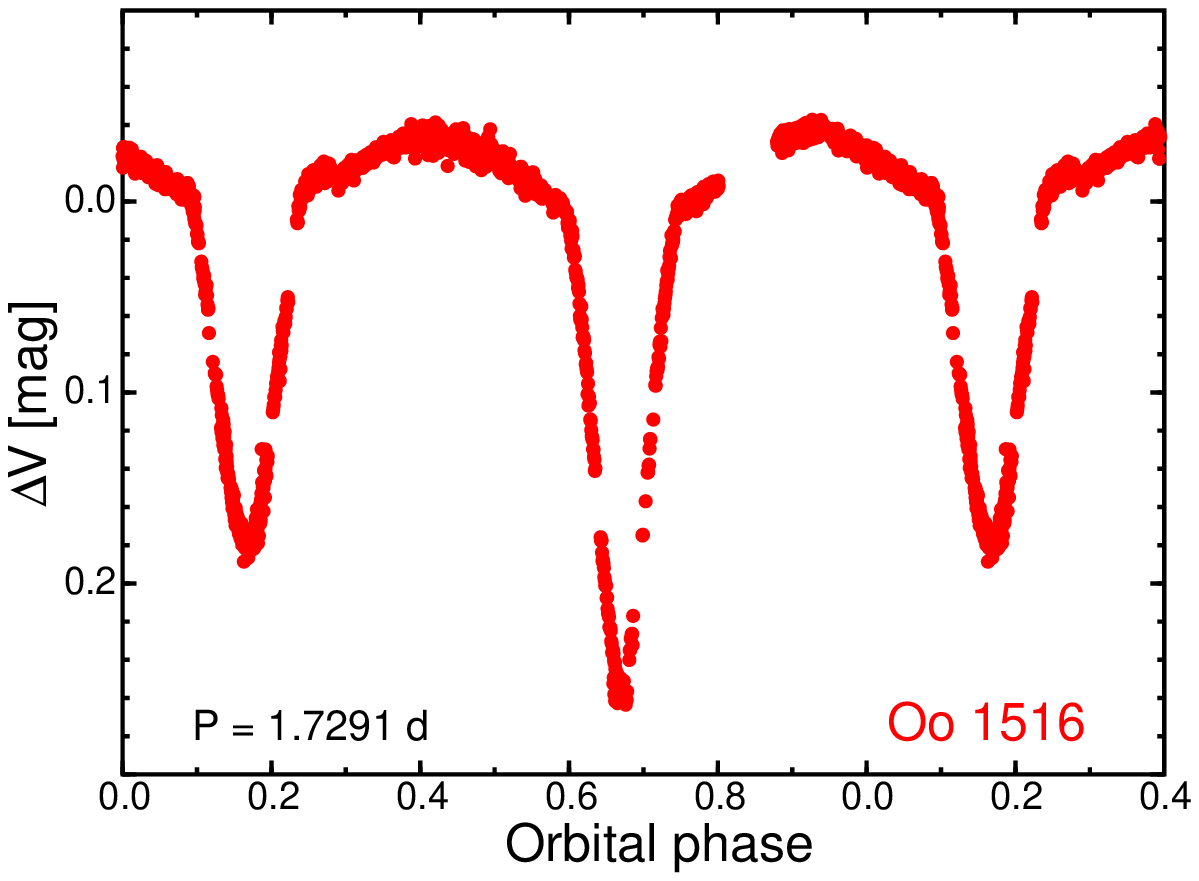}
\end{center}
\caption{\label{ecl}Light variations of the eclipsing variable, Oo 1516, phased with the orbital period of 1.7291 d.}
\end{figure}

\section{The results}
h Persei was already a target of variability search. Krzesi\'nski et al.~\cite{krze99} observed only the central 
part of the cluster, however, finding two $\beta$ Cephei stars, Oo 692 and 992, and several other variables. 
Another $\beta$ Cephei star in the cluster, Oo 839, was found by Gomez-Forrellad \cite{gofo00}. Our prelimary 
analysis shows that h Persei is as rich in $\beta$ Cephei stars as its twin, $\chi$ Persei. In this note
 we confirm the short-period variability in Oo 692, 839 and 992 and announce the discovery of (multi)periodic 
variability in six early B-type members of the cluster (Oo 306, 530, 778, 803, 859 and 1001). These six stars are 
therefore very likely $\beta$~Cephei stars. 

The location of all nine $\beta$~Cephei stars in the colour-magnitude diagram of h Persei is shown in Fig.~\ref{cmd}, 
while their light curves, in Fig.~\ref{BCeps}. Several candidates (not shown) were also found. It is interesting 
to note that in three stars, Oo 530, 692, and 778, periodic variations with periods longer than 0.35 d were found. 
A possible explanation is the $g$-mode variability. In addition, h Persei seems to be very rich in variable 
B-type stars of other types. In particular, we have found over 40 stars which show (quasi)periodic variations 
with periods ranging from 0.3 d up to several days. The variability of these stars is very likely due to 
multiperiodic pulsation, although due to aliasing our data are insufficient to yield accurate frequencies
for some of them. Taking into account the observed periods, these are $g$ modes or other modes that occur 
in the presence of fast rotation (see, e.g., \cite{town05}). Eight stars from this sample are known Be stars, 
i.e., they are fast rotating stars by definition, but some of the remaining stars also rotate fast. In the colour-magnitude 
diagram (Fig.~\ref{cmd}) stars from this group (light and dark green dots) occupy the whole region of B-type 
stars in h Persei. We conclude that most of them may be classified as SPB/SPBe stars. A sample light curves 
showing this kind of variability for two SPB candidates and two Be stars are shown in Fig.~\ref{SPBe}.

In addition, we found three $\delta$ Scuti stars, of which two can be members of the cluster (see Fig.~\ref{cmd}). 
Finally, several eclipsing binaries including three W UMa-type systems were found. The light curve of the 
brightest eclipsing system, Oo 1516, is shown in Fig.~\ref{ecl}. As can be seen from Fig.~\ref{cmd}, W UMa-type 
binaries might be cluster members as well.

\ack{A.P.~is grateful to the EC for the establishment of the European Helio- and Asteroseismology Network
HELAS, which made his participation at this workshop possible. The work was partially supported by the MNiI 
grant No.~1 P03D 016 27.}

\end{document}